\documentclass[a4,aps,prd,amsmath,rotating,twocolumn,preprintnumbers,nofootinbib,floatfix]{revtex4-1} 
\usepackage{color,ulem}
\usepackage{graphicx}
\newcommand{\htext}[1]{{\color{red}  #1}}

\begin{document}

\title{Gravitino cosmology with a very light neutralino}
\author{Herbi~K.~Dreiner}
\email[]{dreiner@th.physik.uni-bonn.de}
\author{Marja~Hanussek}
\email[]{hanussek@th.physik.uni-bonn.de}
\affiliation{Bethe Center for Theoretical Physics and Physikalisches Institut, 
University of Bonn, Bonn, Germany}

\author{Jong~Soo Kim} 
\email[]{jongsoo.kim@tu-dortmund.de}
\affiliation{Institut f\"ur Physik, Technische Universit\"at Dortmund,
  Dortmund, Germany, and\\
ARC Centre of Excellence for
Particle Physics at the Terascale,
School of Chemistry and Physics,
 University of Adelaide, Adelaide, Australia
 }

\author{Subir Sarkar}
\affiliation{Rudolf Peierls Centre for Theoretical Physics, University
  of Oxford, Oxford OX1 3NP, UK}

\begin{abstract}
It has been shown that very light or even massless neutralinos are
consistent with all current experiments, given non-universal gaugino
masses.  Furthermore, a very light neutralino is consistent with
astrophysical bounds from supernov{\ae} and cosmological bounds on
dark matter. Here we study the cosmological constraints on this
scenario from Big Bang nucleosynthesis taking gravitinos into account
and find that a very light neutralino is even favoured by current
observations.
\end{abstract}

\preprint{DO-TH-10/23}
\preprint{ADP-11-36/T758}

\maketitle

\section{Introduction}
Within the minimal supersymmetric Standard Model (MSSM), the photon
and the $Z^0$ boson, as well as the two neutral CP-even Higgs bosons,
have SUSY spin-1/2 partners which mix. The resulting mass eigenstates
are denoted neutralinos, $\chi^0_i$, with $i=1, \ldots,4$, and are
ordered by mass $m_{\chi^0_1}<\ldots<m_ {\chi^0_4}$
\cite{Martin:1997ns}. The Particle Data Group quotes a lower mass
bound on the lightest neutralino \cite{Nakamura:2010zzi}
\begin{equation} 
m_{\chi^0_1} > 46 \textrm{ GeV}\,,
\label{pdg}
\end{equation} 
which is derived from the LEP chargino search under the assumption of
gaugino mass universality:
\begin{equation}
M_1=\frac{5}{3}\tan^2\theta_\text{W} M_2\,.
\end{equation}
Here $M_{1,2}$ are the supersymmetry (SUSY) breaking bino mass and
wino mass, respectively and $\theta_\text{W}$ is the electroweak
mixing angle.  If we relax this latter assumption, the bound
(\ref{pdg}) no longer applies. In fact for any value of $M_2,\,\mu,$
and $\tan\beta$ there is always a $M_1$
\begin{eqnarray}
M_1 &=& \frac{M_2 M_Z^2\sin(2\beta)\sin^2\theta_\text{W}}{\mu
 M_2 - M_Z^2 \sin(2\beta)\cos^2\theta_\text{W}} \label{massless}\\
&\simeq& 2.5 \textrm{ GeV} \left( \frac{10}{\tan\beta} \right)
\left( \frac{150 \textrm{ GeV}}{\mu} \right)\,,
\end{eqnarray}
such that the lightest neutralino is massless
\cite{Gogoladze:2002xp,Dreiner:2009ic}. Here $M_Z$ is the mass of the
$Z^0$ boson, $\tan\beta$ is the ratio of the vacuum expectation values
of the two $CP$--even neutral Higgs bosons in the MSSM and $\mu$ is
the Higgs mixing parameter of the superpotential. A very light or
massless neutralino is necessarily predominantly bino--like since the
experimental lower bound on the chargino mass, sets lower limits on
$M_2$ and $\mu$ \cite{Choudhury:1999tn,Dreiner:2006sb}. Although
Eq.~(\ref{massless}) holds at tree--level, there is always a massless
solution even after including quantum corrections to the neutralino
mass \cite{Dreiner:2009ic}.

Such a light or even massless neutralino is consistent with all
laboratory data.  The processes considered include the invisible width
of the $Z^0$, electroweak precision observables, direct pair
production, associated production, and rare meson decays. Note that a
bino-like neutralino does not couple directly to the $Z^0$. The other
production processes, including the meson decays, thus necessarily
involve virtual sleptons or squarks.  If these have masses of ${\cal
  O}(200)$~GeV or heavier, then all bounds are evaded --- for details
on the individual analyses see
Refs.~\cite{Choudhury:1999tn,Dedes:2001zia,Gogoladze:2002xp,Barger:2005hb,Dreiner:2003wh,Dreiner:2006sb,Dreiner:2009er,Dreiner:2009ic}.
The best possible laboratory mass measurement can be performed at a
linear collider via selectron pair production with an accuracy of
order 1 GeV, depending on the selectron mass \cite{Conley:2010jk}.

Light neutralinos can lead to rapid cooling of supernov{\ae}, so are
constrained by the broad agreement between the expected neutrino pulse
from core collapse and observations of SN~1987A \cite{Ellis:1988aa}.
The neutralinos would be produced and interact via the exchange of
virtual selectrons and squarks.  For a massless neutralino which
`free-streams' out of the supernova, the selectron must be heavier
than about 1.2 TeV and the squarks must be heavier than about 360
GeV. For light selectrons or squarks of mass $\sim 100-300$~GeV, the
neutralinos instead diffuse out of the supernova just as the neutrinos
do and thus play an important role in the supernova dynamics. Hence
lacking a detailed simulation which includes the effects of neutralino
diffusion, no definitive statement can presently be
made~\cite{Ellis:1988aa,Grifols:1988fw,Kachelriess:2000dz,Dreiner:2003wh}.
Recently the luminosity function of white dwarfs has been determined
to high precision \cite{A8Silvestri:2006vz,A8Isern:2008fs} and this
may imply interesting new bounds on light neutralinos, just as on
axions.

If a neutralino is stable on cosmological time scales it can
contribute to the dark matter (DM) of the universe.  If `cold', then
its mass is constrained from below by the usual Lee-Weinberg bound
\cite{Lee:1977ua} which depends only on the self-annihilation
cross-section. This limit has been widely discussed in the literature
in the framework of the $\Lambda$CDM cosmology
\cite{Hooper:2002nq,Belanger:2002nr,Belanger:2003wb,Bottino:2003iu}
and various values are quoted for a MSSM neutralino: $M_{\chi^0_1}>
12.6\,$ GeV \cite{Vasquez:2010ru,arXiv:1108.1338} and $M_{
\chi^0_1}>9\,$ GeV \cite{Fornengo:2010mk,arXiv:1108.2190}. The low
mass range is particularly interesting because the DAMA 
\cite{Bernabei:2010mq} and CoGeNT \cite{Aalseth:2010vx} direct 
detection experiments have presented evidence for annual modulation
signals suggestive of a DM particle with mass of ${\cal O}(10)$~GeV.

A light neutralino with a much smaller mass is also viable as `warm'
or `hot' DM but this possibility has been less discussed. The observed
DM density $\Omega_\text{DM} h^2 \approx 0.11$ can in principle be
entirely accounted for with warm dark matter (WDM) in the form of
neutralinos having a mass of a few keV \cite{Profumo:2008yg}.  However
the usual assumption of radiation domination and entropy conservation
prior to big bang nucleosynthesis (BBN) then needs to be relaxed
otherwise the relic neutralino density is nominally much larger than
required. This scenario requires a (unspecified) late episode of
entropy production or, equivalently, reheating after inflation to a
rather low temperature of a few MeV.  Although models of baryogenesis
with such reheating temperatures
exist~\cite{Allahverdi:2010im,Kohri:2009ka}, the necessary baryon
number violating interactions would result in rapid decay of the
proton to (the lighter) neutralinos. This makes such models very
difficult to realise in this context, although the situation may be
somewhat eased since the maximum temperature during reheating can be
higher than the final thermalisation temperature
\cite{Davidson:2000dw}.

In this paper we focus on a light neutralino which acts as hot dark
matter (HDM)\footnote{Note that HDM cannot contribute more than a
  small fraction of the observed dark matter, so another particle is
   required to make up the cold dark matter (CDM). Potential
  candidates include the gravitino \cite{Ellis:1984er}, the axion
  \cite{Preskill:1982cy} or the axino \cite{Rajagopal:1990yx}.},
\emph{i.e.} can suppress cosmic density fluctuations on small scales
through free-streaming. In order for its relic abundance to be small
enough to be consistent with the observed small-scale structure we
require \cite{Dreiner:2009ic} following Ref.\cite{Cowsik:1972gh}:
\begin{equation} 
m_{\chi^0_1}\lesssim 0.7 \textrm{ eV}\,.
\end{equation}
Such ultralight neutralinos affect BBN by contributing to the
relativistic degrees of freedom and thus speeding up the expansion
rate of the universe; consequently neutron-proton decoupling occurs
earlier and the mass fraction of primordial ${}^4$He is increased
\cite{Sarkar:1995dd}. The resulting constraint on new relativistic
degrees of freedom is usually presented as a limit on the number of
additional effective $SU(2)$ doublet neutrinos:
\begin{equation}
 \Delta N^{\textrm{eff}}_\nu{({\chi^0_1})} 
\equiv N^{{\rm eff}}_\nu-3\,. 
\end{equation}
In \S\,\ref{DeltaNuNeut}, we calculate this number in detail and
compare it with observational bounds on $\Delta N^{\textrm{eff}}_\nu$
from BBN \cite{Cyburt:2004yc}. 

Until recently, the BBN prediction and
the inferred primordial $^4$He abundance implied according to some
authors \cite{Barger2003,Simha:2008zj} 
\begin{equation}
\Delta N^{\rm eff}_\nu\lesssim0\,.
\end{equation}
This is however in tension with recent measurements of the cosmic
microwave background (CMB) anisotropy by WMAP, which suggest a
larger value of \cite{Komatsu:2010fb,Hamann:2010bk}
\begin{equation}
\mathrm{WMAP:}\quad  \Delta N^{\rm eff}_\nu = 1.34\,^{+0.86}_{-0.88}\,.
\label{CMBNeff} 
\end{equation}
Recent measurements of the primordial $^4$He abundance are also higher
than reported earlier, implying \cite{Izotov:2010ca, Aver:2010wq}:
\begin{equation} 
\mathrm{BBN:}\quad  \Delta N^{\rm eff}_{\nu}=0.68\,^{+0.8}_{-0.7} \,.
\label{boundNeff} 
\end{equation} 
Given these large uncertainties, a very light neutralino is easily
accommodated, and even favoured, by the BBN and CMB data.  In the near
future, the Planck mission \cite{Tauber:2011ah} is foreseen to
determine $N^{\rm eff}_\nu$ to a higher precision of about $\delta
N^{\rm eff}_\nu = \pm 0.26$~\cite{Hamann:2010bk}, thus possibly
constraining the light neutralino hypothesis.

Local SUSY models necessarily include a massive gravitino
\cite{Freedman:1976xh}. Depending on its mass, the gravitino can also
contribute to $\Delta N^{\textrm{eff}}_\nu$ as we discuss in
\S\,\ref{DeltaNuNeutGrav}. This effect is only relevant for sub-eV mass
gravitinos (for models see \textit{e.g.}
Ref.~\cite{Brignole:1997pe}).  More commonly the gravitino has
electroweak-scale mass and its decays into the light neutralino will
result in photo-dissociation of light elements, in particular ${}^4$He
\cite{Sarkar:1995dd}. The resulting (over) production of ${}^2$H and
${}^3$He is strongly constrained observationally and we present the
resulting bounds in \S\,\ref{gravitinodecays}.  In 
\S\,\ref{quasistable} we examine under which conditions the gravitino 
itself can be a viable DM candidate in the 
presence of a very light neutralino. Conclusions are presented in 
\S\,\ref{conclusion}.

\section{Light neutralinos and nucleosynthesis}
\label{DeltaNuNeut}

In global SUSY models, or local SUSY models with a non--relativistic
gravitino, the sub--eV neutralino is the only relativistic particle
present at the onset of nucleosynthesis apart from the usual photons,
electrons and 3 types of neutrinos.

The contribution of the neutralino to the number of
effective neutrino species is \cite{Sarkar:1995dd}:
\begin{equation}
\Delta N_{\nu}^{\rm eff}(\chi^0_1) = \frac{g_{{\chi^0_1}}}{2} \left(
\frac{T_{{\chi^0_1}}}{T_{\nu}}\right)^4,
\label{eq:delta_nu}
\end{equation}
where $g_{{\chi^0_1}}$ is the number of internal degrees of freedom,
equal to 2 due to the Majorana character of the neutralino. The ratio
of temperatures is given by
\begin{equation} 
\frac{T_{{\chi^0_1}}}{T_{\nu}} = \left[ \frac{g^* (T_{\rm
     fr}^{\nu})}{g^*(T_{\rm fr}^{{\chi^0_1}})} \right]^{1/3}\,, 
\label{eq:TchiOverTnu}
\end{equation} 
where $T_{\rm{fr}}^i$ is the freeze--out temperature of particle $i$
and
\begin{equation}
g^*(T)=\sum_{\rm{bosons}} g_i\cdot \left(\frac{T_i}{T}\right)^4 + 
\frac{7}{8} \sum_{\rm{fermions}} g_i \cdot\left(\frac{T_i}{T}\right)^4.
\end{equation}
with $g_i$ being the internal relativistic degrees of freedom at
temperature $T$. Usually $T_i$ for a decoupled particle species $i$ is
lower than the photon temperature $T$. because of subsequent entropy
generation.

The freeze-out temperature of $SU(2)$ doublet neutrinos is $T_{\rm
  fr}^{\nu} \sim 2$~MeV \cite{Dicus:1982bz}. The interaction rate
$\Gamma_{\chi^0_1}$ of the lightest neutralino is suppressed relative
to that of neutrinos \cite{Dreiner:2009ic} because the SUSY mass scale
$m_{\rm SUSY} >M_\text{W}$, where $m_{\rm SUSY}$ denotes the relevant
SUSY particle mass involved in the neutralino reactions. Hence the
freeze-out temperature of the very light neutralino will generally be
higher than $T_{\rm fr}^{\nu}$.

Estimating the thermally-averaged neutralino annihilation
cross-section via an effective vertex, we obtain the approximate
interaction rate
\begin{equation}
\Gamma_{\chi^0_1}(T) = 2\frac{3}{4}\frac{\zeta(3)}{\pi^2} G_{\rm
  SUSY}^2\,T_{\chi^0_1}^5,
\end{equation}
where $G_{\rm SUSY}/\sqrt{2}=g^2/(8 m_{\rm SUSY}^2)$. Equating this to
the Hubble expansion rate~\cite{Sarkar:1995dd}
\begin{equation}
  H (T)=\sqrt{\frac{4\pi^3g^*(T)}{45}}\frac{T^2}{M_{\rm{Pl}}} ,
\label{hubble}
\end{equation}
where $g^*$ counts the relativistic degrees of freedom, yields the
approximate freeze-out temperature:
\begin{equation}
T_{\rm fr}^{{\chi^0_1}} \approx 3 \; \left(\frac{m_{\rm SUSY}}{200\textrm{
    GeV}}\right)^{4/3} T_{\rm fr}^{\nu} \,.
\label{eq:TNeutEstimate}
\end{equation}
Thus, for sparticle masses below $\sim 3$ TeV, the 
neutralinos freeze--out below the temperature at which muons
annihilate \cite{Dreiner:2009ic}.

We now calculate the freeze--out temperature of a pure bino--like
neutralino more carefully, considering all annihilation processes into
leptons which are present at the time of neutralino freeze--out:
\begin{equation} 
{\chi^0_1}{\chi^0_1}\rightarrow \ell\bar
  \ell,\quad \ell=e,\nu_e,\nu_\mu,\nu_\tau.
\label{eq:annihilation}
\end{equation}
Assuming that sleptons and sneutrinos have a common mass scale
$m_{\mathrm{slepton}}$, the following relations hold 
\begin{eqnarray}
  \sigma ({\chi^0_1}{\chi^0_1} \rightarrow \ell_R \bar \ell_L) & = &
  16\sigma ({\chi^0_1}{\chi^0_1} \rightarrow \ell_L\bar\ell_R)
  \nonumber\\ & = & 16\sigma({\chi^0_1}{\chi^0_1} \rightarrow
  \nu\bar\nu),
\end{eqnarray} 
so the total annihilation cross section into leptons is given by
\begin{equation} 
\sigma({\chi^0_1}{\chi^0_1}\rightarrow \ell\bar\ell) = 20 \sigma
({\chi^0_1}{\chi^0_1}\rightarrow \ell_L\bar \ell_R)\,,
\end{equation} 
where we have taken the electron to be massless. The
thermally-averaged cross-section is then given by
\begin{equation} 
\langle \sigma({\chi^0_1}{\chi^0_1}\rightarrow \ell\bar
\ell) v\rangle= \frac{20}{9
  \zeta(3)^2}\frac{2^5}{3}I(1)^2\hat\sigma T^2,
\label{eq:thermally_averaged}
\end{equation}
where 
\begin{equation}
I(n)=\int_0^{\infty}\frac{y^{n+2}}{\exp(y)+1}
\end{equation}
and 
\begin{equation}
\hat\sigma=\frac{e^4}{8\pi\cos^4\theta_ W}\frac{1}{
  m_{\mathrm{slepton}}^4}
\end{equation}
for $m_{\rm slepton}\gg T$. In calculating the cross-section
(\ref{eq:thermally_averaged}), we have neglected the Pauli blocking
factors in the final state statistics \cite{Gherghetta:1996fm}.

\begin{figure}[t]
\includegraphics[width=\columnwidth]{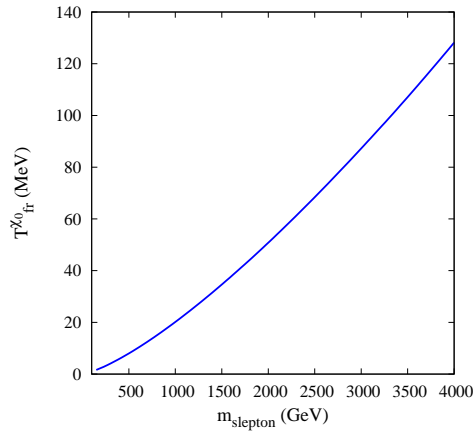}
\caption{Freeze-out temperature of the pure bino--like neutralino as a
  function of the common mass scale $m_{\mathrm{slepton}}$.}
\label{plot:decoupling}
\end{figure}

Relating the reaction rate (\ref{eq:thermally_averaged}) to the Hubble
expansion rate (\ref{hubble}), we can now obtain the freeze--out
temperature for a bino--like neutralino, shown in
Fig.~\ref{plot:decoupling} as a function of the common mass scale
$m_{\mathrm{slepton}}$. Note that for $m_{\mathrm{slepton}}$ below a few TeV, the
neutralino decouples below the muon mass as noted earlier. Thus
neutrinos and neutralinos will have the same temperature,
\begin{equation}
  T_{{\chi^0_1}}=T_{\nu}\,,
\end{equation}
hence during BBN,
\begin{equation} 
\Delta N_\nu^{\rm eff}({\chi^0_1}) = 1\;.
\label{eq:rough_result_neutralino1}
\end{equation} 
However, for slepton masses above a few TeV, the neutralino
freeze--out temperature is close to the muon mass, and muon
annihilation will influence the neutralino and neutrino temperature
differently. For $T^{{\chi^0_1}}_{\rm fr}\gtrsim m_\mu$, the neutrinos
are heated by the muon annihilations, whereas this affects the
neutralinos only marginally. Therefore $T_{{\chi^0_1}}/T_{\nu}$ is
reduced due to the conservation of comoving entropy.  The muons
contribute to $g^*(T_{\chi^0_1})$, such that
\begin{equation}
\frac{T_{{\chi^0_1}}}{T_{\nu}} = \left[ \frac{g_\gamma +
    \frac{7}{8}(\, g_e + 3 g_\nu)}{g_\gamma + \frac{7}{8} (g_e + 3
    g_\nu + g_\mu)} \right]^{1/3} = \left (\frac{43}{57} \right)^{1/3}
.
\end{equation}
Thus employing Eq.~(\ref{eq:delta_nu}) we obtain
\begin{equation} 
\Delta N_\nu^{\rm eff}({\chi^0_1}) = 0.69 \;,
\label{eq:rough_result_neutralino2}
\end{equation} 
which is interestingly close to the observationally inferred central
value of 0.68 in Eq.~(\ref{boundNeff}). The LHC already restricts the
masses of strongly coupled SUSY particles (squarks and gluinos) to be
above several hundred GeV
\cite{daCosta:2011qk,Chatrchyan:2011ek,Bechtle:2011dm} and the
supernova cooling argument requires the selectron mass to also be
above a TeV for a massless neutralino \cite{Dreiner:2003wh}, so the
picture is consistent.

Even for a neutralino freeze--out temperature somewhat below the muon
mass, the effects from muon annihilation are notable. We now determine
the equivalent number of neutrino species more carefully using the
Boltzmann equation as in Refs.~\cite{Dicus:1982bz,Ellis:1985fp}, in
order to determine the effect for arbitrary slepton masses. Consider a
fiducial relativistic fermion $x$ which is decoupled during $\mu
\bar\mu$ annihilation, so that its number density, $n_x$, satisfies
\begin{equation}
\dot n_x+ \frac{3 \dot R}{R} n_x=0\,.
\end{equation}
The Boltzmann equation controlling the number density of the lightest
neutralino can then be written as
\begin{equation}
\frac{d}{dt} \left(\frac{n_{{\chi^0_1}}}{n_x}\right)=n_x\langle\sigma v
\rangle \: \left[\left(\frac{n_\mu}{n_x}\right)^2-f(T_{\chi^0_1})
  \left(\frac{n_{{\chi^0_1}}}{n_x}\right)^2\right],
\label{eq:boltzmann}
\end{equation}
where
\begin{equation}
f (T_{{\chi^0_1}}) =
\left[\frac{n_\mu(T_{{\chi^0_1}})}{n_{{\chi^0_1}}(T_{{\chi^0_1}})}\right
]^2_{\rm{equilibrium}}.
\end{equation}
The cross-section $\mu\bar\mu\rightarrow{\chi^0_1}{\chi^0_1}$ is 
given by
\begin{eqnarray}
16\pi s^2&&\frac{\cos\theta_{\rm
    W}^4}{e^4}\;\sigma(\mu_R\bar\mu_L\rightarrow
{\chi^0_1}{\chi^0_1})=\\ 2(m_{\tilde \mu}^2 -&&m_\mu^2)\,
\ln\left(\frac{2(m_{\tilde \mu}^2
  -m_\mu^2)+s-\sqrt{s}\sqrt{s-4m_\mu^2}}{2(m_{\tilde \mu}^2
  -m_\mu^2)+s+\sqrt{s}\sqrt{s-4m_\mu^2}}\right)\nonumber\\ &&+\sqrt{s}\sqrt{s-4m_\mu^2}\;
\frac{2(m_{\tilde \mu}^2 -m_\mu^2)^2+m_{\tilde \mu}^2s}{(m_{\tilde \mu}^2
  -m_\mu^2)^2+m_{\tilde \mu}^2s} .
\end{eqnarray}
Since this involves a cancellation between the two
terms, we Taylor expand to ensure numerical stability:
\begin{eqnarray}
16\pi\frac{\cos\theta_{\rm W}^4}{e^4}\sigma(\mu_R\bar\mu_L\rightarrow
{\chi^0_1}{\chi^0_1})&\approx&
\frac{\sqrt{1-\frac{4m_\mu^2}{s}}(s-m_\mu^2)}{3(m_{\tilde
    \mu}^2-m_\mu^2)^2}, \nonumber
\end{eqnarray}
then take the thermal average $\langle\sigma v\rangle$ following
Ref.~\cite{Gondolo:1990dk}.

In order to reformulate Eq.~(\ref{eq:boltzmann}) in terms of
dimensionless quantities, we define
\begin{equation}
\delta \equiv \frac{T_{{\chi^0_1}}-T_x}{T_x}, \quad \epsilon \equiv
\frac{T_{\gamma}-T_x}{T_x}\,, \quad y \equiv \frac{m_\mu}{T_\gamma}\,.
\end{equation}
Here $\delta$ measures the temperature difference between the
decoupled particle $x$ and the lightest neutralino and thus quantifies
the heating of the lightest neutralino due to $\mu\bar\mu$
annihilation.  We now evaluate ${n_\mu}/{n_x}$ numerically and expand
${n_ {{\chi^0_1}}}/{n_x}\approx1+3\delta$ so Eq.~(\ref{eq:boltzmann}) can
be written as \cite{Dicus:1982bz,Ellis:1985fp}
\begin{equation}
\frac{d\delta}{dy}\approx ay^{-2}(\epsilon-\delta)\,,
\label{eq:delta-eq}
\end{equation}
for $\delta\ll1$, \textit{i.e.} for small temperature differences.
The prefactor $a$ depends on the size of the annihilation
cross-section, and thus on $y$ and the slepton mass:
\begin{equation}
a(y, m_{\tilde l}) = \frac{ 5.67 \times 10^{17} }{\sqrt{g^*}} \frac{
  \langle \sigma v\rangle }{\textrm{ GeV}^{-2}}\,.
\end{equation}
We approximate the drop in $g^*$ when the muons become
non--relativistic by a step-function with $g^*(y<1)=16$ and
$g^*(y>1) = 12.34$.

\begin{figure}[t]
\includegraphics[width=\columnwidth]{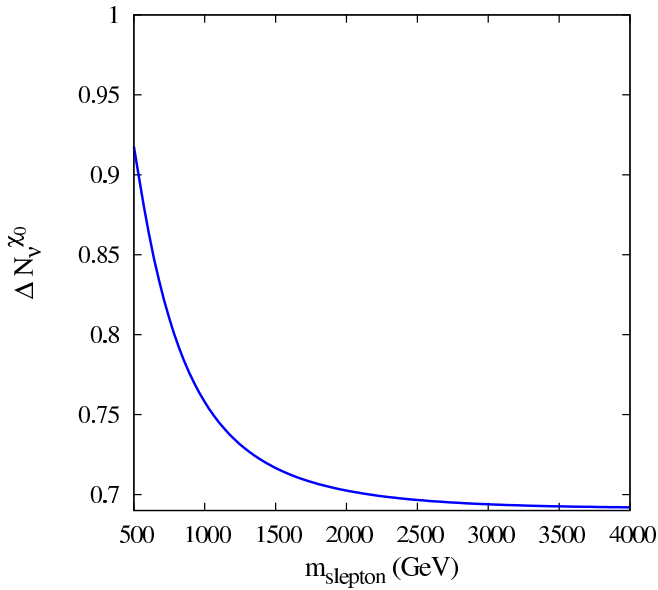}
\caption{Contribution of the pure bino--like neutralino to the
  effective number of neutrinos versus the slepton mass.}
\label{plot:deltanu} 
\end{figure}

Now $T_x$ and the photon temperature $T_\gamma$ are related through
entropy conservation~\cite{Sarkar:1995dd}:
\begin{equation}
\frac{T_x}{T_\gamma}=\left(\frac{43}{57}\right)^{1/3}\left[\zeta(y)
\right]^{1/3},
\label{eq:TxTgamma}
\end{equation}
where
\begin{eqnarray}
\zeta(y)&=&1+ \frac{180}{43\pi^4} \nonumber\\ & & \times
\int_0^{\infty} x^2 \frac{\sqrt{x^2+y^2}+\frac{x^2}{3
    \sqrt{x^2+y^2}}}{e^{\sqrt{x^2+y^2}}+1}\, \text{d}x. \label{eq:zeta}
\end{eqnarray}
We use Eqs.~(\ref{eq:TxTgamma}) and~(\ref{eq:zeta}) to numerically
evaluate $\epsilon(y)$ and then solve the differential equation
(\ref{eq:delta-eq}) for $\delta(y, m_{\tilde l})$.  The solution
asymptotically approaches a limit [denoted by $\delta_{\rm
max}(m_{\tilde l})$] for $y \gtrsim 10$ because for
temperatures far below the muon mass there is no further heating of
the neutralinos from muon annihilation. This improves our estimate
(\ref{eq:rough_result_neutralino1}) to:
\begin{equation}
\Delta N_\nu^{\rm eff}({\chi^0_1})=\left(\frac{T_{{\chi^0_1}}}
{T_\nu}\right)^4
=0.69\,[1+\delta_{\rm max}(m_{\tilde l})]^4.
\end{equation}
In Fig.~\ref{plot:deltanu}, we show $\Delta N_\nu^{\rm
  eff}({\chi^0_1})$ as a function of the common slepton mass $m_{\mathrm{slepton}}$. 
  We see that for slepton masses above 3 TeV, our previous
result of 0.69 in Eq.(\ref{eq:rough_result_neutralino2}) is not
modified. This is because if the interaction between the neutralinos
and muons is too weak, then the neutralinos cannot stay in thermal
contact with the muons.  For slepton masses around 1 TeV, we get again
1 additional effective neutrino species. (Our numerical approximation
is valid only for $\delta\ll1$, so holds down to $m_{\mathrm{slepton}}=0.5$
TeV when $\delta\simeq 0.1$.)

Summarizing, the neutralino contribution to the effective number of
neutrinos lies between 0.69 and 1, depending on the slepton mass as
seen in Fig.~\ref{plot:deltanu}. Thus, a very light neutralino is
easily accommodated by BBN and CMB data and is in fact favoured by the
recent observational indication (\ref{boundNeff}) that $N_\nu \gtrsim
3$ .

\section{A very light neutralino and a very light gravitino}
\label{DeltaNuNeutGrav}

A very light gravitino (as realized \textit{e.g.} in some models of
gauge-mediated SUSY breaking) can constitute HDM. For its relic
density to be small enough to be consistent with the observed
small--scale structure requires \cite{Feng:2010ij}:
\begin{equation}
m_{\tilde G} \lesssim 15 - 30 \textrm{ eV}\,.
\label{gravitino-mass}
\end{equation}
If the gravitino is heavier than the (very light) neutralino it will
decay into it plus a photon with a lifetime $\gtrsim 10^{38}$~s [see
Eq.~(\ref{gravdecay}) below] which is well above the age of
the universe $\sim 4\times10^{17}$~s. Conversely if the gravitino is
lighter than the neutralino, the latter will decay to a gravitino and
a photon with lifetime \cite{Covi:2009bk}
\begin{equation}
\tau_{\chi^0_1} \simeq 7.3 \times 10^{41} \textrm{ s}
\left(\frac{m_{\chi^0_1}}{1 \textrm{ eV}} \right)^{-5}
\left(\frac{m_{\tilde G}}{0.1 \textrm{ eV}} \right)^{2} \,,
\end{equation}
assuming that there is no near--mass degeneracy between the neutralino
and the gravitino. Again the lifetime is well above the age of the
universe, therefore we can consider both the gravitino and the very
light neutralino as effectively stable HDM.

The presence of a very light gravitino thus affects the primordial
$^4$He abundance analogously to a very light neutralino.  However, the
contribution of the gravitino to the expansion rate depends on its
mass, since it couples to other particles predominantly via its
helicity--1/2 components with the coupling strength $\Delta
m^2/(m_{\tilde G} m_{\rm Pl})$, where $\Delta m^2$ is the squared mass
splitting of the superpartners \cite{Fayet:1977vd}. For a very light
gravitino, the interaction cross-section can be of order the weak
interaction, leading to later decoupling. Hence it can have a sizeable
effect on BBN.

The freeze-out temperature of a very light gravitino can be
estimated from the conversion process with
cross-section~\cite{Fayet:1979yb}
\begin{equation}
\sigma(\tilde G e^\pm\rightarrow e^\pm{\chi^0_1})=\frac{\alpha}{9}\frac{s}{m_{\rm
    Pl}^2m_{{\tilde G}}^2}.    
\label{eq:conversion}
\end{equation}
We neglect self--annihilations, ${\tilde G}{\tilde G}\rightarrow\ell
\bar\ell,\gamma\gamma$ since the annihilation rate into photons is
$\propto m_{\chi^0_1}^4$ \cite{Gherghetta:1996fm,Bhattacharya:1988ey} hence
suppressed for a light neutralino, while the annihilation rate into
leptons is $\propto T^6$~\cite{Gherghetta:1996fm} so falls out of
equilibrium much earlier than the conversions.
\begin{figure}[t]
\includegraphics[width=\columnwidth]{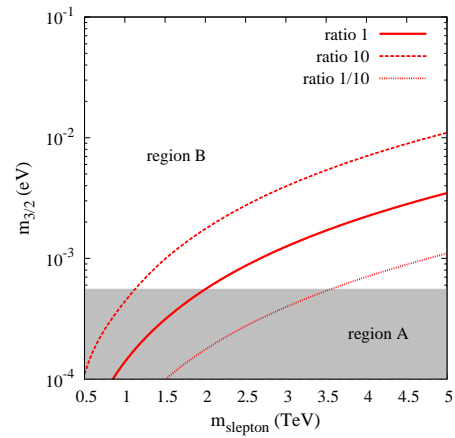}
\caption{Contour lines for the ratio of cross-sections for neutralino
  self--annihilation (\ref{eq:annihilation}) and conversion
  (\ref{eq:conversion}), in the gravitino--slepton mass plane. The
  shaded area indicates where $\Delta N_\nu^ {\rm total}=2$.  }
\label{plot:ratio}
\end{figure}

After thermal averaging of the conversion rate
(\ref{eq:conversion}) as before, we find
\begin{eqnarray}
T_{\rm{fr}}^{\rm conversion}&\simeq& 7.51\, m_{\tilde
  G}^{2/3}m_{\rm{Pl}}^{1/3} g^{*^{1/6}} \nonumber \\ &\approx& 100
\,g^{*^{1/6}} \left(\frac{m_{\tilde G}}{10^{-3} \textrm{ eV}}\right)^{2/3}
\textrm{ MeV}.
\label{eq:gravfreezeout1}
\end{eqnarray}
Since the goldstino coupling is enhanced for decreasing gravitino
mass, the freeze-out temperature of the gravitino increases with its
mass. For a gravitino mass of $5.6\times10^ {-4} $~eV ($7.8\times
10^{-4}$ eV) its freeze-out temperature equals the muon (pion) mass,
so for heavier gravitinos the contribution to $\Delta N_\nu^{\rm eff}$
will decrease.  We also consider the case $m_{\tilde G}=10$ eV which
gives a freeze-out temperature of ${\cal O} (100)$~GeV, thus a
negligible effect on $\Delta N_\nu^{\rm eff}$.  (Note however that
$T_{\rm{fr}}^{\rm {\tilde G}}$ will now depend on the SUSY mass
spectrum because above temperatures of a GeV or so other SUSY
processses can also be in thermal equilibrium
\cite{Fayet:1982gg,Chung:1997rq} and Eq.~(\ref{eq:gravfreezeout1}) may
not apply.)

We can now evaluate the contribution of the gravitino, in conjunction
with the very light neutralino, to the effective number of neutrino
species. We need to keep in mind that the gravitino can affect
neutralino decoupling since for very large slepton masses and/or very
light gravitinos, the neutralino annihilation process
${\chi^0_1}{\chi^0_1}\rightarrow\ell\bar\ell$ becomes sub--dominant to the
conversion process $\tilde G e^\pm\rightarrow e^\pm{\chi^0_1}$ and
therefore neutralino freeze--out is also governed by
Eq.~(\ref{eq:gravfreezeout1}).

In Fig.~\ref{plot:ratio}, we show contour lines for the ratio of the
cross-sections for neutralino annihiliation (\ref{eq:annihilation}),
and the conversion process (\ref{eq:conversion}), in the
slepton--gravitino mass plane. For a ratio less than 0.1, the
freeze--out temperature of both particles is determined via the
conversion process (\ref{eq:conversion}) and $T_{{\tilde G}}=T_{\chi^
0_1}$. Hence $\Delta N_\nu^{\rm eff}({\tilde G},{\chi^0_1})=1/0.69/
0.57$, the latter two cases corresponding to gravitino masses above
$5.6\times10^ {-4} $~eV and $7.8\times 10^{-4}$ eV, respectively
[corresponding to a freeze--out temperature below the muon and
the pion mass, as determined from 
Eq.~(\ref{eq:gravfreezeout1})]. The corresponding equivalent
number of neutrino species is:
\begin{equation} 
\Delta N_\nu^{\rm total}\equiv \Delta N_\nu^{\rm eff}({\tilde G}) + \Delta
N_\nu^{\rm eff}({\chi^0_1})=2/1.38/1.14 \,.
\end{equation} 
Thus a very light gravitino is strongly constrained by the BBN
bound~(\ref{boundNeff}), a mass below $5.6\times10^ {-4} $~eV being
excluded at 3$\sigma$.  As the gravitino mass increases, $\Delta
N_\nu^{\rm total}$ decreases because the gravitino and neutralino
freeze-out earlier, hence are colder than the neutrinos at the onset
of BBN.

One can see from Fig.~\ref{plot:ratio} that a further increase of the
gravitino mass (or smaller slepton mass) accesses parameter regions
where the neutralino annihilation process dominates over the
conversion process. When the ratio of their rates exceeds $\sim 10$,
the freeze--out of the neutralino and the gravitino is governed by the
processes (\ref{eq:annihilation}) and (\ref{eq:conversion})
respectively. For a slepton mass above $\sim 3$ TeV, the lightest
neutralino decouples above the muon mass hence yields $\Delta
N_\nu^{\rm eff}({\chi^0_1})=0.69$. Fig.~\ref{plot:deltanu} shows that with
decreasing slepton mass, this increases to $\Delta N_\nu^{\rm
  eff}({\chi^0_1})=1$ as before. Hence we obtain the same bounds on the
gravitino mass for $\Delta N_\nu^{\rm eff}({\tilde G})=1/0.69/0.57$.
  
In summary for a slepton mass below $\sim1$ TeV 
\begin{equation}
\Delta N_\nu^{\rm total}=2/1.69/ 1.57\, ,
\end{equation} 
while for a slepton mass above $\sim 3$~TeV
\begin{equation}
\Delta N_\nu^
{\rm total}=1.69/1.38/1.26\,;
\end{equation}
for intermediate slepton masses, there is a continuous transition
between the two cases.

If the gravitino mass increases further its effect on the expansion
rate continues to decrease, \textit{e.g.} for $m_{\tilde G} = 10$ eV
(corresponding to $T_{\rm fr}^{\tilde G}\approx100$ GeV), we find $g^*
= 395/4$ or $\Delta N_\nu^{\rm eff}({\tilde G})\simeq 0.05$. Thus,
gravitinos with mass $\gtrsim$ eV do not significantly affect the
expansion rate.

Summarising, $\Delta N_\nu^{\rm total}$ is between 1.14 and 2 for
scenarios with both a relativistic neutralino and a relativistic
gravitino (when their freeze-out temperature lies between the
freeze-out temperature of the neutrino and the pion mass).  As before
we can use the Boltzmann equation if necessary to obtain exact
values for $\Delta N_\nu^{\rm eff}$ around the mass thresholds. From
Eq.~(\ref{boundNeff}), $N_\nu^{\rm total}> 4.9$ is excluded at
$3\sigma$ implying a lower bound on the gravitino mass of
$5.6\times10^{-4}$ eV, \textit{cf.}\ Fig.~\ref{plot:ratio}. This
bound is two orders of magnitude weaker than the one stated in
Ref.~\cite{Gherghetta:1996fm} where a model with a very light
gravitino but a heavy neutralino was considered. This is because the
gravitino annihilation into di-photons or leptons is the relevant
process when there is no light neutralino, also
Ref.~\cite{Gherghetta:1996fm} assumed a more stringent BBN limit:
$N_\nu^{\rm total}<3.6$.

\section{Decaying Gravitinos}\label{gravitinodecays}

So far we have considered the increase in the expansion rate caused by
sub--eV neutralinos and gravitinos which are quasi--stable
(\textit{cf.}  \S\,\ref{quasistable}). We now consider a gravitino
with a mass above ${\cal O}(100\,\mathrm{GeV})$ as would be the case
in gravity mediated SUSY breaking where the gravitino sets the mass
scale of SUSY partners.

As the gravitino mass increases, the relative coupling strength of the
helicity--1/2 components, $\Delta m^2/(m_{\tilde G} m_{\rm Pl})$
decreases and the helicity--3/2 components come to dominate. These are
however also suppressed by 1/m$_{\rm Pl}$ hence gravitinos decouple
from thermal equilibrium very early. During reheating, gravitinos are
produced thermally via two--body scattering processes (dominantly QCD
interactions) and the gravitino abundance is proportional to the
reheating temperature $T_\text{R}$ \cite{Ellis:1984eq}. The gravitino
is unstable and will decay subsequently into the very light neutralino
and a photon with
lifetime~\cite{Ellis:1984eq,Ellis:1984er,Ellis:1990nb,Kawasaki2005},
\begin{equation}
\tau_{\tilde G} \simeq 4.9\times 10^8
\;\left(\frac{m_{3/2}}{100\,\rm{GeV}}\right)^{-3}\rm{s}\,,
\label{gravdecay}
\end{equation}
where we have assumed for simplicity that the gravitino is the
next-to-lightest-SUSY particle (NLSP) while the neutralino is the
lightest-SUSY particle (LSP). If the gravitino decays around or after
BBN, the light element abundances are affected by the decay products
whether photons or hadrons.  In particular there is potential
overproduction of D and $^3$He from photodissociation of (the much
more abundant) $^4$He \cite{Ellis:1984er,Ellis:1990nb}, while for
short lifetimes, decays into hadrons have more effect
\cite{Kawasaki2005}.

Therefore, the observationally inferred light element abundances
constrain the number density of gravitinos. For a gravitino lifetime
of $\mathcal{O}(10^8\,$sec) one obtains \cite{Kohri2006a,Cyburt2009a}
a severe bound on the abundance $Y_{3/2}\equiv n_{3/2}/s$:
\begin{equation} Y_{3/2} \lesssim 10^{-14} \;
\left(\frac{100\,\rm{GeV}}{m_{\tilde G}}\right).  
\end{equation} 
This is proportional to the reheating temperature through
\cite{Ellis:1984eq,Ellis:1984er,Ellis:1990nb,Kawasaki2005}
\begin{equation} 
\left(\frac{T_\text{R}}{10^{10}\,\rm{GeV}} \right) \approx 3.0\times10^{11}
\; Y_{3/2}\,,
\end{equation} 
hence the latter is constrained to be
\begin{equation} 
T_\text{R} \lesssim 3.0\times10^{7} \textrm{ GeV} \times
\left(\frac{100\,\rm{GeV}}{m_{3/2}}\right)\;.
\end{equation} 
Note that a reheating temperature below $\mathcal{O}(10^8$ GeV) is not
consistent with thermal leptogenesis, which typically requires
$T_\text{R}\sim10^{10}$ GeV \cite{Fukugita:1986hr}. There are however
other possible means to produce the baryon asymmetry of the universe
at lower
temperature~\cite{Allahverdi:2010im,Davidson:2000dw,Kohri:2009ka}.

The contribution to the present neutralino relic density from
gravitino decays is
\begin{equation}
\Omega^{\textrm{decay}}_{\chi_1^0}h^2\approx0.28 \; Y_{3/2}\left(
\frac{m_{\chi_1^0}}{1\,\rm{eV}}\right)\,.
\label{eq:upperbounds}
\end{equation}
\textit{i.e.} negligible, such that the Cowsik--McClelland bound on
the neutralino mass is unaffected.

\section{Quasi--stable Gravitinos}\label{quasistable}

As mentioned in \S\,\ref{gravitinodecays}, when the gravitino mass is
below $\sim 100$ MeV its lifetime is longer than the age of the
universe so it is quasi--stable and can constitute warm dark
matter. Decaying gravitino DM is constrained by limits on the diffuse
$\gamma$--ray background. For a mass between $\sim100$ keV and $\sim
100$ MeV the gravitino decays to a photon and a neutralino, and the
photon spectrum is simply
\begin{equation}
\frac{dN_{\gamma}}{dE} = \delta(E - \frac{m_{\tilde G}}{2}) \,.
\end{equation}
The $\gamma$--flux from gravitions decaying in our Milky Way halo
dominates ~\cite{Buchmuller:2007ui,Bertone:2007aw} over the redshifted
flux from gravitino decays at cosmological distances. Using a
Navarro-Frenk-White profile for the distribution of DM in our galaxy,
we obtain
\begin{eqnarray}
&E^2& \frac{{\rm d}J}{{\rm d}E}|_{\rm halo} \equiv \frac{2 E^2}{8 \pi
    \tau_{{\tilde G}} m_{{\tilde G}}} \frac{{\rm d}N_\gamma}{{\rm d}E}
  \int_{\rm l.o.s} \langle \rho_{\rm halo}(\vec\ell) d\vec\ell\,
  \rangle / \Delta\Omega \nonumber\\ &\;&=31.1 \left( \frac{m_{\tilde
      G}}{1 \textrm{ MeV}} \right)^4 \delta(E - \frac{m_{\tilde
      G}}{2}) \frac{\textrm {MeV}}{\textrm{cm}^2 \textrm{ str s}}\,.
\end{eqnarray}
We compare this to the measurements of the $\gamma$-ray background by
COMPTEL, EGRET and Fermi
\cite{Strong:2004ry,Strong:2005zx,Abdo:2010nz} and extract a
conservative upper bound of $3 \times 10^{-2}$
cm$^{-2}$str$^{-1}$s$^{-1}$MeV on the $\gamma$-ray flux from the inner
Galaxy in the relevant mass region below $\sim 100$ MeV. This implies
that gravitinos with mass above $\sim 250$ keV would generate a flux
exceeding the observed galactic $\gamma$-ray emission.  On the other
hand, constraints from small--scale structure formation set a lower
mass bound on WDM of
$\mathcal{O}$(keV)~\cite{Viel:2007mv,deVega:2009ku,Boyarsky:2008ju}.


Now we consider the relic density of those gravitinos.  Due to the
presence of the very light neutralino, all sparticles will decay into
the latter before the onset of BBN. Therefore the gravitino will only
be produced thermally with relic density~\cite{Pradler:2007ne}
\begin{equation}
\Omega_{3/2} h^2 \approx \left(\frac {1 \textrm{ keV}}{m_{\tilde G}}\right)
\left(\frac{T_\text{R}}{10 \textrm{ TeV}}\right) \left(\frac {M_{\rm SUSY}}{200
  \textrm{ GeV}}\right)^2 .
  \label{eq:omega}
\end{equation}
This further restricts the gravitino mass and/or the reheating
temperature in order not to exceed the observed value $\Omega_{\rm DM}
h^2\approx 0.11$.  The least restrictive upper bound on the reheating
temperature from Eq.~(\ref{eq:omega}) is $\mathcal{O}(10^5\,
\mathrm{GeV})$ for gravitino and gaugino masses of order 100 keV and
100 GeV, respectively.  This could be alleviated if the gravitino
density is diluted by the decay of particles (such as moduli
fields~\cite{Kawasaki2005} or the saxion from the axion
multiplet~\cite{Hasenkamp:2010if,Pradler:2006hh}).  In this context,
there have been several detailed studies on gravitinos as light DM
~\cite{Kawasaki:1996wc,Baltz:2001rq,Jedamzik:2005ir,Gorbunov:2008ui,Heckman:2008jy}.

\section{Summary}
\label{conclusion}

We have studied the cosmology of the gravitino in the presence of a
very light neutralino. Even a massless neutralino is compatible with
all laboratory data, while the strictest astrophysical
constraint is imposed by supernova cooling and requires
selectrons to be heavy ($m_{\tilde e}\gtrsim1\,$TeV). Here we have
considered the effect of a stable very light neutralino arising on the
effective number of neutrino species during big bang
nucleosynthesis. For slepton masses above $\sim\!3$ TeV, $\Delta
N_\nu^{\mathrm{eff}}(\chi^0_1)$ is 0.69 and this increases as the
slepton mass decreases, reaching 1 for slepton masses below
$\sim\!0.5$ TeV.

Next, we have considered constraints on the gravitino mass in the
context of local SUSY with a very light neutralino. A very light
gravitino will affect the expansion rate of the universe similarly to
a light neutralino.  We have identified the mass range where a
gravitino has a sizeable effect on the effective number of neutrino
species as $\sim 10^{-4}-10$ eV.  Within this range, we obtain values
for $\Delta N_\nu^{\mathrm{eff}}(\chi^0_1 \,\& \,{\tilde G})$ between
0.74 and 1.69, depending on the gravitino and slepton masses.  Values
around 0.7 are favored by recent BBN measurements. However, the
uncertainties in the determination of ${}^4$He are still sufficiently
large that we need to await data from Planck to pin down the allowed
gravitino and slepton mass.

If the gravitino is heavier than $\sim 100$ MeV, it decays to the
neutralino and a photon with a lifetime smaller than the age of the
universe. This results in photo-dissociation of the light elements,
which is strongly constrained observationally and translates into an
upper bound on the reheating temperature of the universe of $\sim
10^7$ GeV for typical gravity mediated SUSY breaking models. Note that
neither the neutralino nor the gravitino can constitute \htext{the 
complete} dark matter in the scenarios considered so far.

The mass range where the gravitino can constitute warm dark matter is
constrained by bounds from the diffuse $\gamma$-ray background, from
the formation of structure on small-scales, and from the observed DM
abundance, leaving a small window of allowed gravitino mass between 1
and 100 keV for a reheating temperature below $10^5$ GeV.

\begin{acknowledgments}
We thank Manuel Drees and Nicolas Bernal for many useful discussions.
MH and JSK would like to thank the Rudolf Peierls Center of
Theoretical Physics at the University of Oxford for their hospitality.
JSK also thanks the Bethe Center of Theoretical Physics and the
Physikalisches Institut at the University of Bonn for their
hospitality.  This work has been supported in part by the ARC Centre
of Excellence for Particle Physics at the Terascale and by the
Initiative and Networking Fund of the Helmholtz Association, contract
HA-101 (ÒPhysics at the TerascaleÓ), the Deutsche Telekom Stiftung and
the Bonn--Cologne Graduate School.
\newline 
\end{acknowledgments}

\end{document}